
-------------------------------------------------------------

\magnification\magstep 1
\vsize=8.5truein
\hsize=6truein
\voffset=0.5 truecm
\hoffset=1truecm
\baselineskip=22pt

\def\hef{$^4\!$He }
\def\het{$^3\!$He }
\def\hem{$^3\!$He-$^4\!$He }

\def\rr{ {\bf r} }
\def\qq{ {\bf q} }
\def\kk{ {\bf k} }
\def\Sqw{$S(q,\omega)$ }
\def\ia{\AA$^{\!-\!1}$\ }

 { }

\vskip 2 truecm

\centerline{\bf DYNAMIC STRUCTURE FUNCTION IN $^3$He-$^4$He MIXTURES}

\vskip 2 truecm

\centerline{J. Boronat}

\centerline{\it Departament de F\'\i sica i Enginyeria Nuclear, }

\centerline{\it Universitat Polit\`ecnica de Catalunya, Pau Gargallo
5, E-08028 Barcelona, Spain}

\smallskip

\centerline{F. Dalfovo}

\centerline{\it Dipartimento di Fisica, Universit\`a di Trento,
I-38050 Povo, Italy}

\smallskip

\centerline{F. Mazzanti and A. Polls}

\centerline{\it Departament d'Estructura i Constituents de la
Mat\`eria,}

\centerline{\it Universitat de Barcelona, Diagonal 645, E-08028
Barcelona, Spain}

\vskip 2 truecm

\noindent {\bf Abstract.}\ \ {\it Relevant features of the
dynamic  structure function $S(q,\omega)$ in \hem mixtures at
zero temperature  are investigated  starting from known properties of
the ground state. Sum  rules are used to fix rigorous constraints
to the different contributions to $S(q,\omega)$, coming from \het
and \hef elementary excitations, as  well as to explore the role
of the cross term $S^{(3,4)}(q,\omega)$. Both the low-$q$
(phonon-roton \hef excitations and 1p-1h \het excitations)
and high-$q$ (deep inelastic scattering) ranges are discussed.}

\vskip 1.5 truecm


\vfill\eject

\centerline{\bf I. INTRODUCTION}

\bigskip

{}From the theoretical viewpoint a dilute solution of \het in \hef
is a  very appealing system, since it is a mixture of fermions
and bosons, having a small difference in mass and interacting via
the same interatomic potential.  In spite of the simplicity of
this picture, the prediction of static and dynamic properties of
\hem mixtures is not at all simple and many problems are still
open. Starting from the first idea that the \het atoms behave as
a Fermi gas of quasiparticles [1], several  phenomenological
theories have been proposed in the past, based on more or less
refined  effective interactions (see Ref.[2] for a review). The
microscopic approach, based on {\it ab initio} calculations with a
realistic interatomic potential, has proven  quite hard.
Recently, quantitative results for the ground state have been
obtained by   means of variational methods [3,4]. In Ref.[4] the
ground  state properties of the mixture are calculated using the
Aziz potential [5] and a Jastrow-type wave  function, including
also triplet correlations  and  summing up the elementary diagrams.
The results for the radial distribution functions or, equivalently,
for the static structure functions, clarify the role of the different
kind of correlations in the system.

A complete microscopic theory for the dynamic properties of the
mixture,  based on the true many-body Hamiltonian, is still not
available. The lack of such  a theory is particularly unpleasant if
one considers that accurate experimental data on the
dynamic structure function \Sqw are now available, both at relatively
low ($0.5$\ia$<q< 2$\ia) [6] and high momentum transfer [7]. On the
other hand, it would be highly desirable to find a  microscopic
basis for different dynamical theories [8-11], based on various
approximations for the linear response function, in terms of
self-energy  or pseudopotentials.

The aim of the present work is to take a first step in
this direction. The main idea is to extract as much information
as possible  about $S(q,\omega)$, in the limit of zero
temperature, using as input  the ground state  calculations of
Ref.[4]. We make use of two different tools: the sum rule formalism
and the   impulse approximation. The former is applied to
separate and compare the different contributions to the dynamic
structure function coming from \het and \hef density excitations,
as well as from the cross term  $S^{(3,4)}(q,\omega)$.   In
particular, we analyse the moments $m_0, m_1$ and $m_3$, where
$$
m_k (q) = \int \! d\omega \ (\hbar \omega)^k S(q,\omega) \ \ \ .
\eqno (1)
$$
The sum rules contain useful information about the structure of
$S(q,\omega)$ to be used both for the analysis of the
experimental  data and as a test of consistency for theoretical
models. We will  devote special attention to the relative weight
of the density and spin dependent dynamic structure functions
of the \het component, as well as to the cross term
$S^{(3,4)}(q,\omega)$; the role of the latter turns  out to be not
negligible in a relevant range of $q$'s. Finally, we will apply the
sum rule analysis to \Sqw in deep inelastic scattering. The exact sum
rules are compared with the results of  the impulse approximation
(IA), taking the momentum distribution from the calculations of
Ref.[12,13].

\vfill\eject

\centerline{\bf II. SUM RULES}

\bigskip

The sum rule formalism has been extensively used in  quantum
many-body systems. A systematic review of sum rules for density
and particle excitations in liquid \hef is given by Stringari
[14], while recent  calculations  in \het are
reported in Ref.[15]. Sum rules  in \hem mixtures have been used
in the past to explore the role of  long range correlations
[16] and, more recently, to analyse  neutron scattering data
[6]. In the following, we  present the  basic formalism and
 calculate sum rules for the different  components of the
dynamic structure function.

Let us introduce the definition of the density operators
$$
\rho_\qq^{(\alpha)} =  \sum_{j=1}^{N_{\alpha}} e^{i \qq \cdot
\rr_j } \ \ \ , \eqno(2)
$$
where $\alpha=3,4$ and $N_\alpha$ is the number of particles of
$\alpha$-type. The  dynamic structure function for density excitations
at zero temperature is defined as
$$
\eqalign{
S^{(\alpha,\beta)}(q,\omega) = &{1\over 2 \sqrt{N_\alpha
N_\beta}} \sum_n \left[ \langle
0|\rho_\qq^{(\alpha)\dagger}|n\rangle \langle n|
\rho_\qq^{(\beta)}|0\rangle \right. \cr & \left. + \langle
0|\rho_\qq^{(\beta)\dagger} |n\rangle \langle n|
\rho_\qq^{(\alpha)} |0\rangle \right] \delta (\omega -\omega_{n0})
\ \ \ . \cr} \eqno (3)
$$
For \het atoms one also introduces  the spin density  operator [17]
$$
{\bf I_q} = \sum_{j=1}^{N_3} \ {\bf I}_j \ e^{ i \qq \cdot \rr_j}
\ \ \ , \eqno (4)
$$
where ${\bf I}_j$ is the  spin of the $j$-atom, and the spin
dependent dynamic structure function
$$
S^{(3,3)}_I (q,\omega) =  {1 \over N_3 I(I+1)} \sum_n |  \langle
n|  {\bf I_q} | 0 \rangle |^2
\delta(\omega - \omega_{n0}) \ \ \ .
\eqno (5)
$$
The total dynamic structure function is then [6,18]
$$
\eqalign{
S(q,\omega) = & \sigma_4~(1-x)~S^{(4,4)}(q,\omega)
+\sigma_3~x~S^{(3,3)}(q,\omega) \cr
& + \sigma_{3,I}~ x~ S^{(3,3)}_I (q,\omega) +2~
\sigma_{34}~ \sqrt{x(1-x)}~ S^{(3,4)} (q,\omega) \ \ \ , \cr }
\eqno (6)
$$
where $x=N_3/(N_3+N_4)$ is the \het concentration. Inclusion
of the cross sections $\sigma$ makes the
quantity $S(q,\omega)$ directly proportional to the measured
neutron scattering cross section, with energy and wave-vector
transfer $\hbar \omega$ and  $q$. The values $\sigma_4 \!=\!
1.34$, $\sigma_3 \! =\! 4.42$, $\sigma_{3,I}\!=\!1.19$, and
$\sigma_{34}\!=\!2.35$ (in units of barns) are taken from Ref.[6].

Sum rules  are  rigorous relations among energy weighted
integrals of \Sqw and ground state properties.
The moments of the different components of \Sqw are defined in the
following way:
$$
m^{(\alpha,\beta)}_{k,(I)} (q) = \int \! d\omega  \ ( \hbar
\omega)^k \ S^{(\alpha,\beta)}_{(I)} (q,\omega) \ \ \ .
\eqno (7)
$$
Inserting  the definitions of $S^{(\alpha,\beta)}
(q,\omega)$ in Eq.~(7), and using the completeness of the
$|n\rangle$ states, one finds [19] that the moments $m_k$
 can be expressed as ground state mean values of
 the density operator, the spin density operator and the
many body Hamiltonian
$$ H= \sum_{j=1}^{N_3}
{-\hbar^2 \over 2 M_3} \nabla_j^2 + \sum_{j=N_3+1}^{N_3+N_4} {-\hbar^2
\over 2 M_4} \nabla_j^2 +\sum_{i<j}^{N_3+N_4} V(|\rr_i-\rr_j|)
\ \ \ .
\eqno (8)
$$
Simple  expressions can be written  for the lowest $k$ moments,
as it has already been done for pure \hef and \het [14,15]. The main
ingredients are the radial distribution functions and the kinetic
energy per particle in the ground state. In the following, we
discuss the $m_0$, $m_1$ and $m_3$ moments.

\bigskip

\centerline{\bf A.\ \  The $m_0$ moment}

\smallskip

The $m_0$ moment of the dynamic structure functions is
given  by
$$
m^{(\alpha,\beta)}_0 (q) =  (N_\alpha N_\beta)^{-1/2} \langle 0
|\rho_\qq^{(\alpha)\dagger}  \rho_\qq^{(\beta)} |0\rangle
\equiv S^{(\alpha,\beta)} (q) \ \ \ ,
\eqno (9)
$$
where $S^{(\alpha,\beta)}(q)$ are the $(\alpha,\beta)$-components
of the static structure  function. The latter is related to the radial
distribution functions  $g^{(\alpha,\beta)}(r)$ through
$$
S^{(\alpha,\beta)} (q) = \delta_{\alpha \beta} + \left(
{N_\alpha  N_\beta \over {\Omega^2} } \right)^{1/2} \int \! d\rr \
[g^{(\alpha, \beta)} (r) -1]\  e^{i \qq \cdot \rr} \  \ \ ,
\eqno (10)
$$
where $\Omega$ is the volume occupied by the system.
For the spin dependent  component one has
$$
m^{(3,3)}_{0,I} (q) =  {1\over {N_3~I(I+1)}} \langle 0|
{\bf I}_{\qq}^{\dag} \! \cdot
\! {\bf I}_{\qq}  |0 \rangle   \equiv S^{(3,3)}_I (q) \ \ \ ,
\eqno (11)
$$
where $S^{(3,3)}_I (q)$ is the spin dependent static structure
function. The latter can be written in terms of the spin radial
distribution function
$$
S^{(3,3)}_I (q) = 1 + {N_3 \over \Omega} \int \! d\rr \
g_I^{(3,3)} (r) \ e^{ i \qq \cdot \rr } \ \ \ , \eqno (12)
$$
where
$$
g_I^{(3,3)} (r) = g_{\uparrow \uparrow}^{(3,3)} (r) - g_{\uparrow
\downarrow}^{(3,3)}
(r)
\ \ \ . \eqno (13)
$$
In Eq.(13), \   $g_{\uparrow \uparrow}^{(3,3)}$ and $g_{\uparrow
\downarrow}^{(3,3)}$ are the distribution functions for \het atoms
having parallel and antiparallel spins respectively, normalized in such
a way that
$g^{(3,3)} (r) = g_{\uparrow \uparrow}^{(3,3)} (r) + g_{\uparrow
\downarrow}^{(3,3)}
(r)$.
All these quantities have been  calculated in
Ref.~[4], where  the  concentration dependence of the
static structure  functions is discussed in detail. In Fig.~1,
we show the typical behaviour of the $m_0$ moments at zero pressure
and for two values of concentration.
The static structure function associated with
the \hef component is nearly the same as in the pure \hef phase. The
\het component behaves almost as a gas of free fermions.
Due to the weakness of spin correlations between \het atoms in the mixture,
the spin dependent  moment is very close to the density moment. In
particular, for $x=0.01$ are  indistinguishable.
Finally, the moment $m_0$ of the cross term $S^{(3,4)}$  is significantly
different from zero for $q$ less than approximately $3$\ia. Its oscillating
behaviour reflects  the fact  that $S^{(3,4)}(q,\omega)$ is not  positively
defined.

\bigskip

\centerline{\bf B.\ \ The $m_1$ moment}

\smallskip

The case of the $m_1$ moment is particularly simple. It
corresponds to the energy weighted sum rule, which can be
expressed in the form of a double commutator, as follows:
$$
m_1^{(\alpha,\beta)} (q) = {1\over {2~(N_{\alpha}N_{\beta})^{1/2}}}
\langle
0|
[[
\rho^{(\alpha)\dagger}_\qq, H],\rho_\qq^{(\beta)}] |0 \rangle
\eqno (14)
$$
and
$$
m_{1,I}^{(3,3)} (q) = {1\over {2~N_3~I(I+1)} } \langle 0|
[[{\bf I}_{\qq}^{\dagger} , H ], {\bf I}_{\qq}] | 0 \rangle \ \ \  .
\eqno (15)
$$
The commutators can be explicitly carried out, using the
Hamiltonian (8), and the results are the
well known f-sum rules:
$$
\eqalignno{
m_1^{(4,4)} (q) & = {\hbar^2 q^2 \over 2 M_4} & (16) \cr
m_{1,I}^{(3,3)} (q)  = m_1^{(3,3)} (q) & = {\hbar^2 q^2 \over 2 M_3}
& (17) \cr
m_1^{(3,4)} (q) &= 0 \ \ \ . & (18) \cr }
$$
The last result supplies  a rather strong model
independent constraint to the cross term $S^{(3,4)}(q,\omega)$.

\bigskip

\centerline{\bf C. \ \  The $m_3$ moment}

\smallskip

The generalization of the calculation of $m_3$ for the pure
phase  [14,15,20] to the case of \hem mixtures is
straightforward. Again, one  has to evaluate commutators between
$\rho_\qq$, ${\bf I}_\qq$ and $H$. The  final  results are:
$$
\eqalign{
m_3^{(4,4)} (q) = & \  {\hbar^6 q^6 \over 8 M_4^3} +
{\hbar^4 q^4 \over M_4^2} t_4 \cr
&  + {\hbar^4 \rho_4 \over 2 M_4^2} \int \! d\rr \  g^{(4,4)}
(r) [1-\cos(\qq \cdot \rr)] (\qq \cdot \nabla)^2 V(r) \cr }
\eqno (19.a)
$$
$$
\eqalign{
m_3^{(3,3)} (q) = & \ {\hbar^6 q^6 \over 8 M_3^3} +
{\hbar^4 q^4 \over M_3^2} t_3 \cr
&  + {\hbar^4 \rho_3 \over 2 M_3^2} \int \! d\rr \  g^{(3,3)}
(r) [1-\cos(\qq \cdot \rr)] (\qq \cdot \nabla)^2 V(r)  \cr }
\eqno (19.b)
$$
$$
\eqalign{
m_{3,I}^{(3,3)} (q) = & \ {\hbar^6 q^6 \over 8 M_3^3} +
{\hbar^4 q^4 \over M_3^2} t_3 \cr
& + {\hbar^4 \rho_3 \over 2 M_3^2} \int \! d\rr \ [g^{(3,3)}(r) -
g^{(3,3)}_I (r) \cos(\qq \cdot \rr)] (\qq \cdot
\nabla)^2 V(r) \cr }
\eqno (19.c)
$$
$$
m_3^{(3,4)} (q) = -{\hbar^4 (\rho_4 \rho_3)^{1/2} \over 2 M_4
M_3 } \int \! d\rr \ g^{(3,4)} (r) \cos(\qq \cdot \rr) (\qq
\cdot \nabla)^2 V(r) \ \ \ ,
\eqno (19.d)
$$
where $t_\alpha$ and $\rho_\alpha$ are the kinetic energy per
particle and the particle density of the $\alpha$ isotope,
respectively. We calculate the moments (19a-d) using the ground
state  results of
Ref.~[4]. In Fig.~2, we show the results for the $m_3$ moments of
the \hef and \het components  at zero pressure  and for
$x=0.06$. The solid curves are the total moments (19.a) and
(19.b), while the dashed lines are the corresponding potential
part, i.e., the terms with the integrals of the interparticle
potential $V(r)$ in the same equations. The spin dependent moment
(19.c) is indistinguishable from the density moment (19.b) on the
scale considered. The oscillating behaviour of the potential parts
reflects the general structure of the radial
distribution functions and the shape of the potential.
 The same holds for the cross moment
(19.d), which is plotted in Fig.3.

\vskip 2 truecm

\centerline{\bf III. \ DENSITY AND SPIN DEPENDENT \het EXCITATIONS}

\bigskip

First of all, we use the above sum rules to study the relative
weight $m_k^{(3,3)}/m_{k,I}^{(3,3)}$
of the density and spin dependent \het excitations.
The case $k\!=\!0$ has been previously discussed in Sec.~II.A. The
density and spin dependent static structure functions are very close
to each other, so that the ratio $(m_0^{(3,3)}/m_{0,I}^{(3,3)})$ is almost
$1$, even at low $q$.  As it is evident from Eq.~(17) the ratio of the first
moments is always equal to $1$. Finally, the ratio
$m_3^{(3,3)}/m_{3,I}^{(3,3)}$  at zero pressure and for two values of
the concentration is plotted in Fig.~4. When all the above ratios are
close to 1, the effects of spin correlations on the dynamic structure
function are negligible, and
$$
S^{(3,3)} (q,\omega) \simeq S^{(3,3)}_I (q,\omega)
\ \ \ . \eqno (20)
$$
This is certainly true for wavevectors much larger than the Fermi
wavevector $q_F$.  At $x=0.06$ the latter turns out to be $0.34$\ia.
We note  that for typical  values of momentum transfer in neutron scattering
experiments one lies well above $q_F$, so that the
difference between the density  and spin dependent structure
functions can be safely neglected.  As an example, we consider the
values
$q=1$\ia and $x=0.045$, taken from a set of experimental values of
Ref.~[6]. With these $q$ and $x$ we find $m_0/m_{0,I}=0.966$,
$m_3/m_{3,I}=1.044$. These numbers give  a microscopic basis to the
approximation (20) used in  the analysis of the experimental
spectrum [6].

\vskip 2 truecm

\centerline{\bf IV. THE CROSS TERM ${\bf S^{(3,4)} (q,\omega)}$ }

\bigskip

The three sum rules (9),(18) and (19.d) provide rigorous
constraints to the cross term $S^{(3,4)} (q,\omega)$. From them
it is easy to show how much $S^{(3,4)} (q,\omega)$ contributes to
the moments of the total dynamic structure function. Let us  take
the four terms at the r.h.s. of Eq.(6); one can calculate their
$k=0$ and $k=3$ moments (including $\sigma$-factors and
concentration) and divide them by the corresponding moments of
the l.h.s.. In Fig.~5 we show the  results for  $x=0.06$ and
zero pressure. It appears that  $S^{(3,4)} (q,\omega)$
gives a small  contribution to the sum rules for $q$
greater  than approximately $3$\ia, while  its contribution
at lower $q$'s is not at all negligible.

Let us focus our attention on the range $1$\ia$<q<1.5$\ia,
where the neutron scattering cross section shows two well
separated peaks, corresponding to \het and \hef excitations [6].
A puzzling point in the experimental data is
a strong quenching of the measured particle-hole peak of \het
excitations with respect to a Fermi quasiparticle model. The data
were  analysed with the assumption $S^{(3,4)}(q,\omega)=0$. Fig.~5
shows  that, in this range of $q$'s, the contributions of the cross
term to the $m_0$ and $m_3$ moments  of the total dynamic structure
function are comparable in magnitude with the ones of
$S^{(3,3)}(q,\omega)$. In particular, $m_0^{(3,4)}$ is negative and
$m_3^{(3,4)}$ is positive, while $m_1^{(3,4)}$ is zero.  This fact
suggests that $S^{(3,4)}(q,\omega)$  should be included in the
analysis of the experimental spectrum. In principle, it  is
possible to satisfy the above $m_k$ moments with  a cross
structure function distributed  at high energies, with negligible
contributions on the  \het and \hef peaks.  This seems, however,
quite unlikely. It is more reasonable to think that
$S^{(3,4)}(q,\omega)$ should have  a  significant part of its
strength in the  region of \het and \hef elementary excitations. To
support this idea we suggest the following model.

Let us suppose that the total dynamic form factor is the sum of:

\item{i)} a peak centered around
$\omega_{\rm p-h}= \hbar^2 q^2/2M_3^*$, where $M_3^*$
is the effective mass of \het quasiparticles. The peak is the sum of
$S^{(3,3)}(q,\omega)$ and $S^{(3,4)}(q,\omega)$. The latter is
negative  and follows approximately the shape of the \het
peak.

\item{ii)} A narrow peak at the frequency $\omega_0(q)$
of the phonon-roton branch, corresponding almost entirely to
$S^{(4,4)}(q,\omega)$.

\item{iii)} A band of multiparticle excitations above $\omega_0$,
corresponding to a
broad distribution of strength coming mainly from $S^{(4,4)}(q,\omega)$
and partly from $S^{(3,3)}(q,\omega)$ and $S^{(3,4)}(q,\omega)$;
the latter being positive and having the same broad distribution
as the other two.

\noindent
A qualitative sketch of the model is shown in Fig.~6. With these
assumptions, and with approximation (20),
it is possible to estimate how much strength  the
\het peak (p-h) takes from the $S^{(3,3)}$ and $S^{(3,4)}$
dynamic structure functions. A semiquantitative analysis is reported
in Table~I for $x=0.045$, and for two values of $q$. The first
three rows are the exact moments calculated in Sec.~II for the cross
structure  function.  Due to the $(\hbar \omega)^3$ factor the $m_3$
moment, as in pure \het and \hef, is almost completely  exhausted
by multiparticle excitations. The ratios $\hbar \omega_{10}= m_1/m_0$
and $\hbar \omega_{31}=\sqrt{m_3/m_1}$ for multiparticle
excitations are directly related to both the average energy
and the broadness of
multiparticle band. As an estimate of these ratios we take their
values  in pure \hef at the same density [21], since the general
structure of the multiparticle band should not be changed
dramatically by the presence of \het. With these numbers we calculate
the multiparticle contributions to the  moments $m^{(3,4)}_0$
and $m^{(3,4)}_1$. The difference between them and the total
sum rules  is an estimate of the strength in the region of the
p-h peak. Now, let us define the dynamic structure function
of the p-h peak as
$$
S_ {\rm p-h} (q,\omega) = S^{(3,3)}_{\rm p-h} (q,\omega)
+  {2 \sigma_{34} \over \sigma_3 + \sigma_{3,I} }
\left( {1-x \over x} \right)^{1\over2} S^{(3,4)}_{\rm p-h} (q,\omega)
\ \ \ . \eqno (21)
$$
For a free Fermi gas $S_{\rm p-h}(q,\omega)$ is a Lindhard function with
integrated strength $S_{\rm p-h} (q) =1$. The p-h peak has been
measured in Ref.~[6] and, through a Lindhard fit, results for the zero- and
first-moments of $S_{\rm p-h}$ have been given.
{}From them, and from the microscopic results for the moments of the
cross term, it is easy to obtain  the moments of
$S^{(3,3)}_{\rm p-h}(q,\omega)$. The  zero-moment turns out to be
almost $1$ and the first moment is reduced by
a factor $M^*_3/M_3 \simeq 3$ with respect to the total f-sum
rule $\hbar^2 q^2/(2M_3)$. Both results are consistent with the
Fermi quasiparticle picture. The possible error in the last
quantities due to the uncertainties in $\omega_{10}$ and
$\omega_{31}$ is found to be  of the order of  $10$\%.
Moreover, we have checked that, within the same accuracy, the
results are concentration independent. The final
outcome of this model is that the strength missing in the measured p-h
peak can be associated with a negative cross term $S^{(3,4)}(q,\omega)$.

These results  are consistent  with the  microscopic sum rules and
with general physical arguments. Even if they should be
considered qualitative rather than quantitative,  they
strongly support the importance of including the cross term in the
theoretical investigation of the dynamic structure function.
The role of the cross term was also studied by L\"ucke and
Szprynger [10] in a RPA-like scheme. In their theory $S^{(3,4)}
(q,\omega)$  is relevant only when the \het quasiparticle peak
and the \hef phonon-roton peak are close and, eventually, overlap.
In this case it produces an asymmetry of the roton peak, which, however,
has not been observed. Our  sum rule analysis seems to justify the
experimental findings; in fact, the contributions of the  cross
term to both the sum  rules $m_0$ and  $m_3$ in Fig.~5  are very
small in the region of overlap between the \het and \hef excitations,
changing  sign near the roton wavelength ($q \simeq 2$\ia).

\vskip 2 truecm

\centerline{\bf V. ${\bf S(q,\omega)}$ IN DEEP INELASTIC SCATTERING}

\bigskip

The density and spin dependent dynamic structure functions,
as defined in Eqs.~(3) and (5), can  always be decomposed in two
pieces [22-24]:
$$
S^{(\alpha,\beta)}(q,\omega)=S^{(\alpha,\beta)}_c
(q,\omega)+S^{(\alpha,\beta)}_{\rm inc} (q,\omega).
\eqno(22)
$$
\noindent The  first term is the coherent structure function and
represents fluctuations which involve different atoms. The second
term is the incoherent structure function (also known as
self-correlation scattering function), which accounts for
correlations between the position of the same atom at different
times. One notes immediately that  the cross term
$S^{(3,4)}(q,\omega)$ is entirely coherent  by definition.
On the other hand, since the interatomic potential is spin independent,
spin correlations between \het atoms arise only from symmetry properties
of the many body wave function; as a consequence,
differences between the density and the spin dependent structure
functions for the \het component are restricted to their coherent parts,
which for $q$ not much larger than $q_F$ become negligible.

The notation of coherent and incoherent scattering is useful in
studying the dynamic structure function at high momentum transfer.
In fact, for sufficiently large $q$,
the wavelength associated to the momentum transfer is small compared
with the mean distance between the scattering atoms. In this  limit,
the piece of $S^{(\alpha,\beta)}(q,\omega)$ due to the interference
of scattering amplitudes  from different atoms is negligible and only
the incoherent $S_{\rm inc}(q,\omega)$ will be left.  Sum rules for the
incoherent
scattering can be easily evaluated; one has to carry out the same
calculations of ground state mean values as in Sec.~II.A-C, but
taking only the $i=j$ terms in the density and spin density  operators.
One finds [22-24]
$$
m_{0,\rm inc}^{(\alpha,\alpha)} (q) =1 \eqno (23.a)
$$
$$
m_{1,\rm inc}^{(\alpha,\alpha)} (q) =  {\hbar^2 q^2 \over 2
M_{\alpha}} \eqno (23.b)
$$
and
$$
m_{3,\rm inc}^{(\alpha,\alpha)}(q) = {\hbar^6 q^6 \over{8
M_\alpha^3}}+ {\hbar^4 q^4 \over {M^2_\alpha}} t_\alpha +
{\hbar^4 \rho_{\alpha} \over {2 M^2_\alpha}} \int \! d \rr \
g^{(\alpha,\alpha)}(r) \ (\qq \cdot \nabla)^2\  V(r) \eqno(23.c)
$$
Notice that $S_{\rm inc}^{(\alpha,\alpha)}$ saturates the total
sum rule $m_1^{(\alpha,\alpha)}$ and therefore the coherent
contribution to the f-sum rule is zero.

Furthermore, one can generalize to the mixture the expansion of
$S_{\rm inc}$ in inverse powers of the momentum transfer $q$,
developed previously for pure phases [22]. The first term of this
expansion is known as the impulse approximation (IA).
In the limit $q \rightarrow \infty$ the scattering simplifies
considerably and only the first term (IA) survives.
Then, the total $S(q,\omega)$ is given by
$$
\lim_{q \rightarrow \infty}S(q,\omega)=S_{IA}(q,\omega)=\sigma_4\,
(1-x)\, S_{IA}^{(4,4)}(q,\omega)+(\sigma_3+\sigma_{3,I})~x~\
S_{IA}^{(3,3)}(q,\omega)     \eqno(24)
$$
where $S_{IA}^{(\alpha,\alpha)}$ are directly related to the atomic
momentum distributions $n_{\alpha}(k)$:
$$
S_{IA}^{(\alpha,\alpha)}(q,\omega)={{ \nu_{\alpha} \hbar}\over {
(2\pi)^3
\rho_{\alpha}}} \int d \kk~ n_{\alpha} (k)~\delta (\hbar \omega-
(\epsilon(\mid \qq+\kk \mid)-\epsilon(k))) ,\eqno(25)
$$
where $\kk$ is the initial momentum of an $\alpha$-atom, $\qq+\kk$
is the final momentum of the recoiling $\alpha$-atom, and
$\epsilon(k)=\hbar^2 k^2/(2 M_{\alpha})$ is its kinetic energy. The
$\delta$ function takes care of the conservation of energy in the
scattering of a neutron from a single atom and $\nu_{\alpha}$ stands
for the spin degeneracy of each component ($\nu_4=1$, $\nu_3=2$).

To calculate $S_{IA}^{(4,4)}$ one should take into account the
macroscopic  occupation of the zero momentum state by  \hef
atoms. To this end, it is convenient to write the momentum
distribution of the different components in the following way:
$$
n_{\alpha}(k)=\delta_{\alpha  4}(2\pi)^{3}\rho_{4}~
n_{0}~\delta(\vec{k}) + \tilde n_{\alpha}(k) \ \ \  ,\eqno(26)
$$
where $n_0$ is the condensate fraction of the \hef component in
the mixture, $\tilde n_{4}(k)$ stands for the  occupation of the
non-zero momentum states of \hef and $\tilde n_{3}(k)$ is the
whole momentum distribution of \het. In the following, we will
write indistinctly $n_{3}(k)$ or $\tilde n_{3}(k)$.  The momentum
distributions $n_{\alpha}(k)$ defined in Eq.~(26) are normalized in
the following way:
$$ {{\nu_{\alpha}}\over {(2 \pi)^3 \rho_{\alpha} }}
\int d {\bf k} ~n_{\alpha}(k)~=~1 \ \ \ . \eqno(27)
$$
Introducing the expression (26) in Eq.~(25) and performing the
corresponding integrations one gets
$$
S_{IA}^{(4,4)}(q,\omega)=n_{0}~\hbar \delta\left(
\hbar \omega-{{\hbar^2~q^2}\over {2~M_4}} \right) + {{M_{4}}\over
{4 \pi^2 \rho_{4}~\hbar~q}} \int_{k_{4}^{min}}
^{\infty} k\tilde n_4(k) dk \ \ \ , \eqno(28.a)
$$
and
$$
S_{IA}^{(3,3)}(q,\omega) = {{M_{3}}\over {2\pi^{2}\rho_{3}\hbar~q}}
\int_{ k_{3}^{min}}^{\infty} kn_3(k)\ dk\ \ \ , \eqno(28.b)
$$
in which
$$
k_{\alpha}^{min}= {{M_{\alpha}}\over {\hbar^2~q}} \left\vert
\hbar \omega
-{{\hbar^2 q^2}\over {2 M_{\alpha}}} \right\vert \ \ \ .  \eqno(29)
$$
Finally, one can also write the expressions for the first sum rules
in impulse approximation:
$$
 m_{0,IA}^{(\alpha, \alpha)}(q)=1+{{\nu_{\alpha}}\over {4\pi^2
\rho_{\alpha}}} \int_{q /2}^{\infty} k \left [{q \over 2}-k
\right ] \tilde n_{\alpha}(k)~ dk \eqno (30.a)
$$
$$
m_{1,IA}^{(\alpha,\alpha)}(q)= {{\hbar^2 q^2} \over {2
M_{\alpha}}}+ {{\nu_{\alpha} \hbar^2 q} \over {8 \pi^2
\rho_{\alpha}M_{\alpha}}} \int_{q/2}^{\infty} k \left [ {q \over 2}-k
\right ]^2  \tilde n_{\alpha} ~dk \eqno (30.b)
$$
$$
 m_{2,IA}^{(\alpha,\alpha)}(q)=\left ( {{\hbar^2 q^2 }\over {2
M_{\alpha}}} \right )^2 + {4 \over 3}{{\hbar^2 q^2}\over {2
M_{\alpha}}}~t_{\alpha} +{{\nu_{\alpha} \hbar^4 q^2}
\over {12 \pi^2 \rho_{\alpha} M_{\alpha}^2}} \int_{q/2}^{\infty} k
\left [ {q  \over 2}- k \right ]^3 \tilde n_{\alpha}(k)~dk  \eqno
(30.c)
$$
$$
m_{3,IA}^{(\alpha,\alpha)}(q)=\left ( {{\hbar^2 q^2}\over {2
M_{\alpha}}}\right )^3+ {{\hbar^4 q^4}\over {M_{\alpha}^2}} t_{\alpha}+
{{\nu_{\alpha}
 \hbar^6 q^3} \over {16 \pi^2 \rho_{\alpha} M_{\alpha}^3}} \int _{q
/2}^{\infty} k  \left [ {q \over 2} -k \right]^4 \tilde n_{\alpha}(k)~
 dk \ \ \ , \eqno (30.d)
$$
The main ingredients to calculate the impulse approximation are the
momentum
distribution of \hef and \het atoms. They have been obtained in the
HNC/FHNC theory from a variational wave function containing two and
three body correlations [12-13].

First, we compare  the exact sum rules, given in
Sec.~II, with the incoherent (23) and the
 impulse approximation (30) sum rules. We
consider the case $x=0.06$ at zero pressure. In
Fig.~7,  we report the moment $m_0(q)$ for \het and \hef excitations.
While $m_{0,\rm inc}^{(4,4)}$ and $m_{0,\rm inc}^{(3,3)}$ are both
equal to one, large differences appear between the total
$m_0^{(3,3)}$ and $m_0^{(4,4)}$. This fact emphasizes
the different
behaviour of the coherent contributions due to the different partial
densities of both components. Since the \het density is very
small, the coherent part of $m_0^{(3,3)}$ goes to zero faster than the
one of $m_0^{(4,4)}$.  Fig.~8 shows the ratio of
$m_{1,IA}^{(3,3)}(q)$ to $m_1^{(3,3)}(q)$. As it has been mentioned
before, $m_1^{(3,3)}$ is entirely incoherent,  and
$m_{1,IA}^{(3,3)}$ reaches the asymptotic limit already at low values
of $q$. The same conclusion applies to the \hef component.  Finally,
we note that the $m_3$ sum rule, except in  the $q \to 0$
limit,
is dominated by  its incoherent part. To proof this statement, the
coherent contribution of $m_3$ for
\hef and
\het excitations  is shown in Fig.~9. The oscillations of
$m_3^{(\alpha,\alpha)}(q)/q^2$
have a similar structure to the one of the total $m_3$ sum rule
of the cross structure function $S^{(3,4)}(q,\omega)$. The results shown
in Figs.~7-9 provide a quantitative information about the quality of
both the impulse and the incoherent approximations at a given
$q$.  Recent experiments of neutron scattering involve momentum
transfer up to $q \simeq 20$\ia or even larger. In this range of
$q$,
the impulse approximation saturates all the sum rules here considered.

Second, we give explicit results for the dynamic structure function in
impulse approximation.  In Fig.~10 we show  \Sqw  calculated using
Eq.(28) for two values of $q$ (15 and 23 \AA$^{-1}$) , and for $x=0.06$.
The deep inelastic peak of each component is well distinguished and
located at  the recoiling energies of each  component, i.e.,
$\hbar \omega_{\alpha}= \hbar^2 q^2/(2~M_{\alpha})$. Notice that the
$\delta$-contribution of the \hef component (see  Eq.~(28.a)) is not
plotted in the figure.
Due to its lighter mass, the peak of the \het is located at higher
energy than that of the \hef, while for low momentum transfer $q$ the
situation is just the opposite, so there should be a range of momenta
where they closely overlap.
The intensity of the \het
peak is substantially smaller than the one of the \hef
because of the low \het concentration.
Nevertheless,
the difference is a little bit compensated by the fact that the \het
cross section for the individual processes is four times larger than
the \hef one. In IA, the height of the peaks decreases inversely
proportional  to $q$ and the distance between them increases roughly
as $\hbar^2 q^2/6 M_{4}$. On the other hand, they
become wider making the overlap of the tails of the responses
not negligible.

The momentum distributions $n_{4}(k)$ and $n_{3}(k)$  have  two
important features. The former has the $k=0$ state occupied by a
fraction of atoms $n_0$, which produces a delta-peak in
$S_{IA}^{(4,4)}(q,\omega)$ at $\hbar \omega = \hbar^2 q^2 / 2 M_4$, not
shown in Fig.~10. On the other hand, $n_3(k)$ has a  gap at
$k_F$, which defines the strength of the quasi-particle pole $Z_F$
at the Fermi surface  and produces a discontinuity in the slope  of
$S_{IA}^{(3,3)}(q,\omega)$  at  $\hbar \omega = \hbar^2 q k_F/M_3
{}~+~\hbar^2~ q^2
/2 M_3$. However, these two features are actually smoothed out by
both
the experimental resolution [25] and the effects of the so called final
state interactions (FSI) [24,26].

Although the experimental uncertainty in $S(q,\omega)$
makes a direct determination of the shape of $n_{\alpha}(k)$ rather
difficult, one can still extract some averaged
quantities. One of them is the kinetic energy per particle.  In
general, the experimental determination of the kinetic energy is
based on the evaluation of the second energy moment of the response
at $q$ large enough so that only the incoherent part contributes. In
that case one has
$$
m_{2,\rm
inc}^{(\alpha,\alpha)}(q)={\left({\hbar^2q^2 \over
{2M_\alpha}}\right)}^2 + {4 \over 3}{\hbar^2 q^2 \over {2 M_\alpha}}
t_\alpha \eqno(31)
$$
Usually, a gaussian centered at
the recoiling energy is fitted to the experimental response and then
one can analitically evaluate the kinetic energy.  While this
procedure gives results for pure \hef [25] in good
agreement with  sophisticated microscopic calculations [27], large
discrepancies appear in the case of pure \het [25,28]. The same
problem appears in the analysis of the mixtures. In fact,  preliminary
experimental results [7] assign to the \het component a kinetic
energy much lower than the theoretical predictions [12,13]. The
latter can be derived  either using the momentum distributions
$n_{\alpha}(k)$ or evaluating the expectation value of the kinetic
energy operator in the trial  variational wave function used to
describe the mixture. For $x=0.06$ and zero pressure the theoretical
values are
approximately $19$ and $14$ K for \het and \hef atoms respectively.
The \hef kinetic energy is almost the same as in pure liquid \hef.
This follows simply from the fact that the density  of the mixture is
only slightly smaller than the saturation density of pure liquid
\hef.
The microscopic calculation of pure \het  at the saturation density
predicts a kinetic energy around $13$ K [28]. The increment in the
kinetic energy of the \het  component in the mixture respect to the pure
phase, is due to the fact that the total density of the mixture is
larger than the pure \het saturation density. Therefore, due to
the
correlations with the \hef atoms, there are more \het  atoms promoted
above the Fermi surface.

We clarify this point in the context of the impulse approximation.
The mean kinetic energy per \het atom $t_3$ can be explicitly
calculated, using the momentum distribution $n_3(k)$, either by
means of a direct integration
$$
t_3 = {{\nu_3} \over {(2\pi)^3 \rho_3}} \int \! d {\bf k} \ {{\hbar^2
k^2} \over {2 M_3}} \  n_3 (k)  \ \ \ , \eqno (32)
$$
or through the definition of $S^{(3,3)}_{IA} (q,\omega)$, as
$$
t_3 = {{3 M_{\alpha}} \over { q^2}} \int_{\omega_3}^\infty
\! d\omega \ (\omega- \omega_3 )^2
S_{IA}^{(3,3)}(q,\omega) \ \ \ , \eqno (33)
$$
where $\hbar \omega_3 = \hbar^2 q^2 / (2 M_3)$. The last equation is
valid  for any momentum transfer, but it is useful from the
experimental point of view only at $q$
large enough so that $S_{IA}^{(3,3)} \sim S^{(3,3)}$. On the other hand,
when $q \rightarrow \infty$ this way to calculate the kinetic energy
is equivalent to the evaluation of $m_{2,\rm inc}^{(3,3)}$, as given
in Eq.~(31). In  order to show the importance of the
high energy tail, we report in Fig.~11 the length of the
energy interval, as a function of $q$, where the integration in
Eq.(33)  should be performed to get $50 \%$ and $95 \%$
of the total $t_3$. For the sake of comparison, it is also plotted
the length of the energy interval needed to integrate the response of
the underlying free Fermi sea (a gas of fermions having the same
$q_F$ as \het atoms in the mixture) to get its total kinetic energy.
The integration interval
grows almost linearly with the momentum transfer. The slope of
the line increases with the percentage of the kinetic energy
required. In the case of the Fermi gas, the interval goes
linearly with $q$ with a well defined slope : $\hbar^2 q_F/M_3$.
Thereby,
one need to reach high momenta in order the IA to be valid but at the
same time, the energy region where it is necessary to
know the response increases with the momentum transfer.
Our results, based on a microscopic calculation of the momentum
distribution, reveal  the  role of the high energy tail of
the \het inelastic peak in determining $t_3$.
Similar conclusions are obtained for the \hef component but, in this
case, the gaussian fit to the deep-inelastic peak leads to a
kinetic energy in good agreement with the microscopic calculations. This
indicates that the gaussian fit can account approximately for the
wings of $S(q,\omega)$. Obviously, this fit is a better
approximation for \hef than for $^3\!$He. The reason is that the
underlying momentum distribution implied by a gaussian
$S^{(\alpha,\alpha)}$ is also of gaussian type and it is clear that this
shape of the momentum distribution reproduces better $n_4(k)$ than
$n_3(k)$. It is also possible to appreciate, by a simple inspection of
Fig. 10, that due to the discontinuity in the slope of $S^{(3,3)}$ it is
more difficult to fit a gaussian to the response of the \het component.
In any case, to compare theory
and experiments one has to take also into account FSI effects, which can
change significatively the structure of the tail  and smooth
out the  discontinuity in the  slope at the Fermi surface, even at large
$q$.

\vskip 2 truecm

\centerline{\bf VI. \ \ CONCLUSIONS}

\bigskip

In this work we have used the sum rule formalism to investigate
properties of the dynamic structure function in \het-\hef mixtures at
zero temperature. The relevant quantities entering the sum rules
have been taken from recent  calculations of the ground state, based
on variational wave functions with two and three body correlations
[4,12,13]. We have discussed the $m_0$, $m_1$ and $m_3$ moments of
the dynamic structure function, separating the different
contributions coming from the \het and \hef components, as well as
from the cross term $S^{(3,4)}(q,\omega)$. The main results can be
summarized as follows:

\item{i)} The role of spin correlations has been investigated. The
differences between the density sum rules and the spin dependent sum
rules  become rapidly small by increasing $q$. The approximation
$S^{(3,3)}(q,\omega)=S^{(3,3)}_I (q,\omega)$ is found to be good
for typical $q$'s used in neutron scattering experiments.

\item{ii)} The cross term $S^{(3,4)}(q,\omega)$ gives contributions
to the total sum rules of the same order than $S^{(3,3)}(q, \omega)$
in  a relevant range of $q$ [29]. A simple qualitative model has been
proposed for the cross term, consistent with the microscopic sum
rules and in agreement with the quenching of the \het p-h peak in the
experimental spectrum [6].

\item{iii)} The dynamic structure function in the deep inelastic
limit has been analysed. The exact sum rules have been compared
with the ones from both the incoherent and the
impulse approximations. A prediction for $S(q,\omega)$ in impulse
approximation has been given. The evaluation  of the mean kinetic
energy per \het particle from the dynamic structure function has been
also discussed.

\bigskip

\noindent {\bf Acknowledgements}

\noindent
This work has been supported in part by DGICYT (Spain) Grant No.
PB89-0332, and by INFN, Gruppo Collegato di Trento.

\vfill\eject

\centerline{\bf REFERENCES}

\bigskip

\item{1.} L.D. Landau and I. Pomeranchuk, Dokl. Akad. Nauk.
SSSR {\bf  59}, 669 (1948).

\item{2.} G. Baym and C. Pethick, in {\it The Physics os Liquid
and  Solid Helium}, part II, K.H. Bennemann and J.B. Ketterson,
eds. (Wiley, New York 1978) p.123.

\item{3.} M. Saarela and E. Krotscheck, in {\it Recent Progress in
Many Body Theories}, Vol. III, pag. 207, Eds. T.L. Ainsworth {\it et al.}
(Plenum Press, New York 1992).

\item{4.} J. Boronat, A. Polls and A. Fabrocini, J. Low Temp. Phys.
in press.

\item{5.} R.A. Aziz, V.P.S. Nain, J.S. Carley, W.L. Taylor, and
G.T.McConville, J. Chem. Phys. {\bf 70}, 4330 (1979).

\item{6.} B. F\aa k, K. Guckelsberger, M. K\"orfer, R. Scherm,
and A.J. Dianoux, Phys. Rev. B {\bf  41}, 8732 (1990).

\item{7.} P.E. Sokol, private communication.

\item{8.} J. Ruvalds, J. Slinkman, A.K. Rajagopal, and A. Bagchi,
Phys. Rev. B {\bf 16},2047 (1977).

\item{9.} K.S. Pedersen and R.A. Cowley, J. Phys. C: Solid
State Phys. {\bf 16} 2671 (1983).

\item{10.} M. L\"ucke and A. Szprynger, Phys. Rev. B {\bf   26},
1374 (1982); A. Szprynger and M. L\"ucke, {\it ibid} {\bf 32},
4442 (1985).

\item{11.} W. Hsu, D. Pines and C.H. Aldrich III, Phys. Rev B
{\bf  32}, 7179 (1985).

\item {12.} J. Boronat, A. Polls, and A. Fabrocini in {\it Condensed
Matter Theories}, Vol. 5, edited by  V.C. Aguilera-Navarro,
(Plenum Press, New York 1990), p.27.

\item {13.} J. Boronat, A. Polls, and A. Fabrocini in preparation.

\item{14.} S. Stringari, Phys. Rev. B {\bf 46}, 2974 (1992).

\item{15.} F. Dalfovo and S. Stringari, Phys. Rev. Lett.  {\bf
63}, 532 (1989).

\item{16.} C.E. Campbell, J. Low Temp. Phys. {\bf 4}, 433
(1971); Hing-Tat Tan, Chia-Wei Woo and F.Y. Wu, {\it ibid} {\bf
5}, 261 (1971).

\item{17.} V.F. Sears, J. Phys. C: Solid State Phys. {\bf 9}, 409
(1976).

\item{18.} H.R. Glyde and E.C. Svensson, in {\it Neutron Scattering},
Vol. 23B of {\it Methods of Experimental Physics}, Edited by D.L.
Price and K. Sk\"old (Academic, New York, 1987),p. 303.

\item{19.} E. Feenberg, in {\it Theory of Quantum Fluids},
(Academic Press, New York and London, 1969).

\item{20.} R.D. Puff, Phys. Rev. A {\bf 137}, 406 (1965).

\item{21.} In pure \hef the multiparticle (or multiphonon)  part
of $S(q,\omega)$ can be extracted from the experimental spectrum
(see, for instance, R.A.Cowley and A.D.B. Woods, Can. J. Phys.
{\bf 49}, 177 (1971)). Although the strength of the multiparticle
band increases with pressure, the average energies $\hbar
\omega_{10}$  and $\hbar \omega_{31}$ are expected to be almost pressure
independent.

\item{22.} H.A. Gersch, L.J. Rodriguez, and P.N. Smith, Phys. Rev.
A  {\bf 5}, 1547 (1972).

\item{23.} S.W.Lovesey, in {\it Theory of Neutron Scattering from
Condensed Matter}, (Oxford Univ. Press, Oxford and New York, 1984);
D.L.Price and K.Sk\"old, in {\it Neutron Scattering}, Vol.23A of
{\it Methods of Experimental Physics}, edited by K.Sk\"old and
D.L.Price (Academic, New York, 1986),p.1.

\item{24.} T.R. Sosnick, W.M. Snow, R.N. Silver, and P.E. Sokol, Phys.
Rev. B {\bf 43}, 216 (1991).

\item {25.} P.E. Sokol, Can. J. Phys. {\bf 65}, 1393 (1987).

\item{26.} C.Carraro and S.E.Koonin, Phys. Rev. Lett. {\bf 65}, 2792
(1990).

\item {27.} M.H. Kalos, M.A. Lee, P. Whitlock, and G.V. Chester, Phys.
Rev. B {\bf 24}, 115 (1981).

\item{28.} R.M. Panoff and J. Carlson , Phys. Rev. Lett. {\bf
62}, 1130 (1989).

\item{29.} A similar conclusion about the non negligible role of
$S^{(3,4)} (q,\omega)$ in the p-h range is also contained in a
very recent work by E. Krotscheck and M. Saarela (unpublished),
based on the RPA formalism.

\vfill\eject

\centerline{\bf FIGURE CAPTIONS}

\vskip 2 truecm

\noindent {\bf Fig.1} : \
The $m_0$ sum rules in the mixture at zero
pressure and for two values of the $^3\!$He concentration.
Solid line: $m_0^{(4,4)}$; long dashed: $m_0^{(3,4)}$;
short dashed: $m_0^{(3,3)}$; dot-dashed: $m_0^{(3,3)I}$.

\bigskip

\noindent {\bf Fig.2} : \
The $m_3$ sum rules in the mixture at zero
pressure and for $x=0.06$. Solid lines: sum rules
$m_3^{(4,4)}$ and $m_3^{(3,3)}$, as in Eqs.(19.a) and (19.b);
dashed lines: potential contribution to the same sum rules,
i.e., the third term in r.h.s. of Eqs.(19.a) and (19.b). The
spin dependent sum rule $m_{3,I}^{(3,3)}$ is practically
indistinguishable from $m_3^{(3,3)}$ on this scale.

\bigskip

\noindent {\bf Fig.3} : \
The $m_3$ sum rule for the cross dynamic structure function
in the mixture at zero pressure and $x=0.06$, as given in Eq.(19.d).

\bigskip

\noindent  {\bf Fig.4} : \
Ratio between the contributions of density and
spin excitations to the $m_3$ sum rule  in
the mixture at two different concentrations and at zero
pressure.  Solid line: $x=0.06$; dashed line: $x=0.01$.

\bigskip

\noindent {\bf Fig.5} : \
Relative contributions of the different
$(\alpha, \beta)$-components to the $m_0$ and $m_3$ moments of
the total dynamic structure function in Eq.(6). The symbol
$\hat m$ means that moments $m_k^{(\alpha,\beta)}$ have been
multiplied by the corresponding $\sigma$ and concentration
factors. Solid line: $(4,4)$; short dashed: $(3,4)$;
long dashed: $(3,3)$; dot-dashed: spin dependent $(3,3)$.
All curves correspond to zero pressure and concentration
$x=0.06$.

\bigskip

\noindent {\bf Fig.6} : \
Qualitative picture for the dynamic structure function in the range $q
\simeq 1$ - $1.5$ \AA$^{-1}$. Solid line: total \Sqw; dotted line:
$S^{(3,3)}$ part of \Sqw in the particle-hole peak; dashed line:
$S^{(3,4)}$ part in the same range. The different contributions to
the multiparticle band are not shown. Scales are arbitrary. Only the
proportions between  the different contributions to the p-h peak are
significant.

\bigskip

\noindent {\bf Fig.7} : \
Moments $m_0^{(\alpha,\alpha)}$ at $x=0.06$. Solid lines: static
structure function; long dashed lines: incoherent approximation;
short dashed lines: impulse approximation.

\bigskip

\noindent {\bf Fig.8} : \
Ratio between the moment $m_1^{(3,3)}$ and the f-sum rule at $x=0.06$.
Solid line: impulse approximation; dashed line: incoherent
approximation.

\bigskip

\noindent {\bf Fig.9} : \
Coherent part of the moments $m_3^{(\alpha,\alpha)}$ divided by
$q^2$, at $x=0.06$. Solid line: $\alpha=4$; dashed line: $\alpha=3$.

\bigskip

\noindent {\bf Fig.10} : \
Dynamic structure function in impulse approximation for two values
of $q$, at zero pressure and $x=0.06$.

\bigskip

\noindent {\bf Fig.11} : \
Length of the energy interval in Eq.(33)
needed  to get 95\% (solid line) and 50\% (long dashed line) of the mean
kinetic
energy $t_3$ as given in Eq.(32). Short dashed line: free Fermi gas.

\vfill\eject

\centerline{\bf TABLE CAPTION}

\bigskip

\noindent {\bf Table I} : \
Quantities entering the analysis of Sec.IV, for
$x=0.045$. First three rows: sum rules calculated in Sec.II.
Rows 4 and 5: sum rule ratios for multiparticle excitations, from
estimates in pure \hef [21]. Rows 6 and 7: $m_0$
and $m_1$ moments of $S^{(3,4)}(q,\omega)$ in the p-h range,
estimated by substracting the multiparticle part from the
exact sum rules. Row 8: measured integrated strength of the p-h
peak (from Fig.~11 of Ref.~[6]). Last two rows:
results for the $m_0$ and $m_1$  moments of $S^{(3,3)}_{\rm
p-h} (q,\omega)$. See the text for a detailed discussion.

\vfill\eject

{}

\vskip 3truecm

\settabs 4\columns
\+& {\bf TABLE I}&\cr

\vskip 4truecm

\+ q \ \ \ [\AA$^{-1}$] & 1.0 & 1.3 &\cr

\smallskip

\hrule
\hrule

\smallskip

\+ $m_0^{(3,4)}$                      & -~0.160    & -~0.125    &\cr
\+ $m_1^{(3,4)}$ \ \ \ [K]            & 0          & 0          &\cr
\+ $m_3^{(3,4)}$ \ \ \ [K$^3$]        & 870        & 3320       &\cr

\smallskip
\hrule
\smallskip

\+ $\hbar \omega_{31}$ \ \ \ [K]            & 60         & 65         &\cr
\+ $\hbar \omega_{10}$ \ \ \ [K]            & 35         & 35         &\cr

\smallskip
\hrule
\smallskip

\+ $(m_0)^{(3,4)}_{\rm p-h}$           &-~0.167     & -~0.148   &\cr
\+ $(m_1)^{(3,4)}_{\rm p-h}$ [K]       &-~0.24      & -~0.79     &\cr

\smallskip
\hrule
\smallskip

\+ $S_{\rm p-h}(q)$        & 0.38       & 0.30       &\cr

\smallskip
\hrule
\smallskip

\+ $(m_0)^{(3,3)}_{\rm p-h}$           & 1.0         & 0.9        &\cr
\+ $(m_1)^{(3,3)}_{\rm p-h}$ [K]       & 2.6         & 4.5        &\cr

\smallskip
\hrule

\bye